\documentclass[10pt, conference, letterpaper]{IEEEtran}

\IEEEoverridecommandlockouts
\usepackage{cite}
\usepackage{amsmath,amssymb,amsfonts}
\usepackage{graphicx}
\usepackage{textcomp}
\usepackage{xcolor}
\usepackage{enumerate}
\usepackage{amsmath}
\usepackage{algorithm}
\usepackage{algpseudocode}
\usepackage{subcaption}
\usepackage{balance}
\def\BibTeX{{\rm B\kern-.05em{\sc i\kern-.025em b}\kern-.08em
    T\kern-.1667em\lower.7ex\hbox{E}\kern-.125emX}}

\begin{document}

\title{Towards High Throughput Wireless Network with Directional Antenna}

\author{Wei Sun\\
The Ohio State University\\
 sun.1868@osu.edu}

\maketitle

\begin{abstract}

In indoor areas such as homes and offices, high throughput communication for multiple devices is quickly becoming a necessity. Even though an access point (AP) mounted with an omni-directional antenna can cover a whole room, it cannot provide connections with high throughput throughout the room. Therefore, we propose, \emph DiRF, a directional antenna based wireless home network designed to achieve high throughput in indoor areas with densely deployed directional APs. \emph DiRF consists of a position based AP selection algorithm which can decrease latency accumulation caused by frequent AP switching, and a downlink packet scheduler which can reduce the downlink packet retransmissions during AP switching. We implement and evaluate \emph DiRF with six commercial APs, each connects with a single directional antenna. Our experiments show that \emph DiRF achieves a $3.16\times$ TCP throughput improvement, compared to the conventional scheme that only uses one AP mounted with one omni-directional antenna.

\end{abstract}


\maketitle

\section{Introduction}

In recent years, it has become common to have multiple wireless devices such as smartphones, laptops, and audio/video equipments that connect with the same WiFi network. These densely deployed WiFi devices compete for access to the wireless medium~\cite{mahajan2006analyzing}, which can reduce the overall network throughput. Also, as the development of high quality applications such as cordless virtual reality (VR)~\cite{abari2017enabling}, augmented reality (AR)~\cite{azuma1997survey} and uncompressed video streaming~\cite{choi2016link}, we require the high capacity  communication between the mobile client and the remote content server for graph rendering and data offloading.

Traditionally, we just deploy one AP with an omni-directional antenna to cover the whole place in a room or an office. This is because AP mounted with omni-directional antenna can provide service to all the WiFi users in the room. However, the omni-directional antenna radiates the radio wave power across all angles uniformly. Therefore, the receiver may not get high signal-to-noise ratio (SNR) at every place of the room during movement, which will degrade the network throughput. Millimeter wave~\cite{wei2017pose} and visible light communication~\cite{komine2004fundamental} can provide high throughput communication by providing high SNR, but they are not widely deployed yet. DIRC~\cite{liu2009dirc} and Speed~\cite{liu2010pushing} measure the optimal orientation of antenna and use a MAC protocol to maximize the transmission concurrency, which cannot account for the throughput degradation during handoff. The existing approaches~\cite{navda2007mobisteer,amiri2010directional} needs more modifications at the mobile user side, which constrains them to be widely used.

\begin{figure}
\centering
    \includegraphics[width=\linewidth]{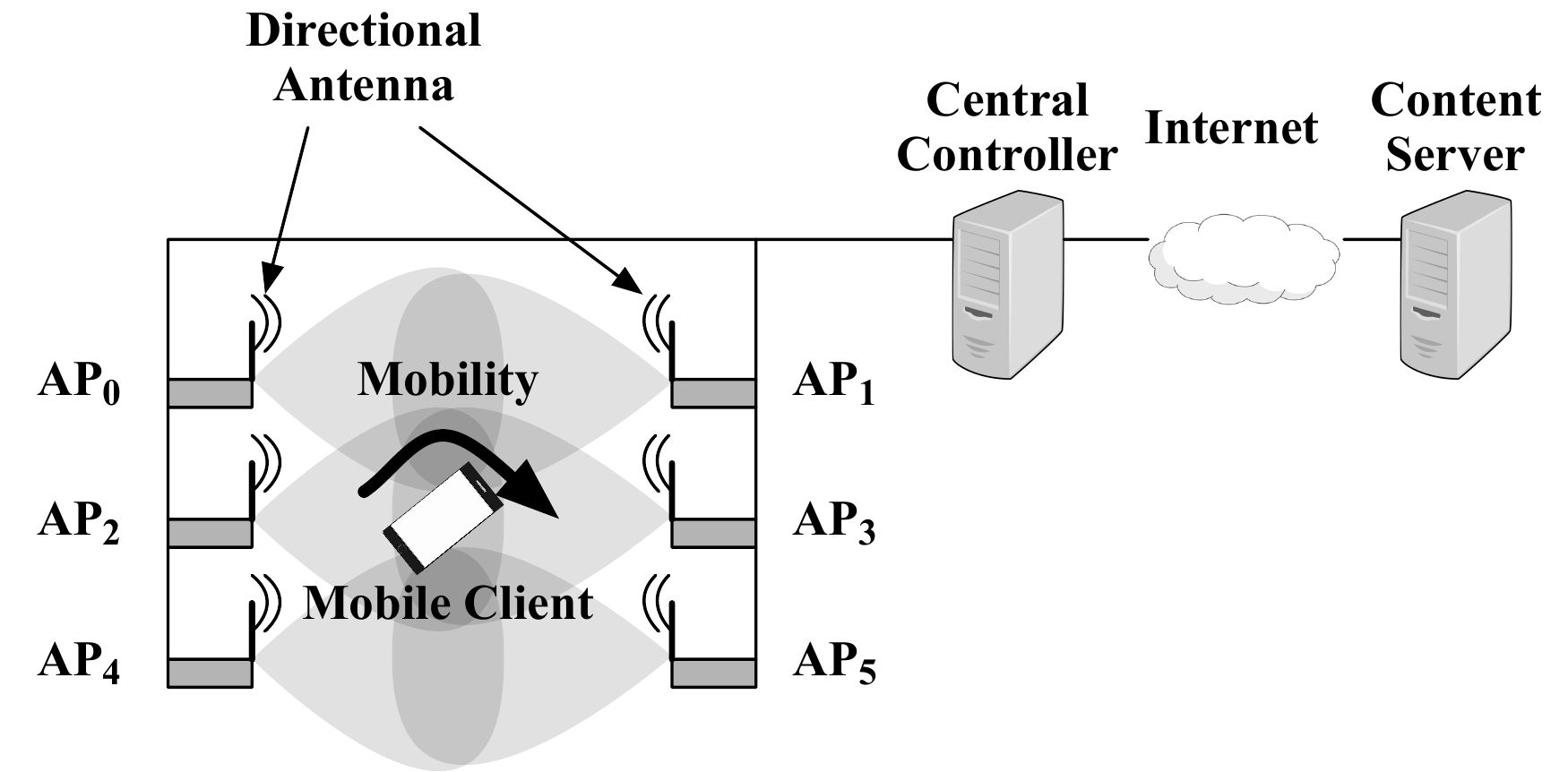}
    \caption{\emph DiRF Architecture: The mobile client moves in each directional antenna enabled AP's coverage. All APs are connected with a central controller through Ethernet cables. The central controller connects the content server through the Internet.}
    \label{fig:arc}
\end{figure}

In this paper, we propose \emph DiRF, a directional AP based wireless network in Fig.~\ref{fig:arc}, which can provide high capacity communication in the indoor environment with densely deployed APs. We deploy multiple APs mounted with the directional antenna to achieve the high capacity communication in an office room. All APs are connected with a content server and a central controller through Ethernet cables. As a mobile client moves in the room it falls into at least one AP's field of view (FoV). Therefore, the mobile client can achieve high network throughput due to the high gain directional antenna, when the mobile client moves in one AP's FoV. When the mobile client goes through the overlap of multiple APs' FoVs, the handoff process is triggered, which degrades the network throughput due to AP switching. Since the network throughput is improved when the client falls in the associated AP's FoV due to high gain directional antenna, it is important to improve the network throughput during AP switching.

To this end, it needs to make a decision on which AP to connect with during handoff. This problem has been investigated extensively in WiFi and cellular networks~\cite{croitoru2015towards, song2017wi, eriksson2008cabernet, soroush2011concurrent, kandula2008fatvap}. Existing AP selection approaches are based on Channel State Information (CSI)~\cite{caire1999capacity}. ClientMarshal~\cite{bhartiaclientmarshal} uses AP's load balancing as a metric to do AP switching. However, simply connecting to the AP with the best CSI and considering AP's load balancing can cause the mobile client to change APs frequently as it continues to move. Roaming among multiple APs has been standardized by 802.11r standard~\cite{4573292} and 802.11k standard~\cite{4544755}, which can allow make-before-break. Also, multipath TCP (MPTCP)~\cite{croitoru2015towards} allows multiple connections between the transmitter and the receiver simultaneously with multiple network interfaces. However, these protocols have not been widely available at the most commercial mobile phones. Also, these standard protocols cannot decrease the retransmissions caused by AP switching.

To achieve high throughput during AP switching, \emph DiRF proposes a novel AP selection algorithm to decrease latency accumulation during movement in the indoor area. Therefore, \emph DiRF leverages the mobile client's  position information extracted from the mobile client's embedded sensors to estimate the mobile client's moving direction. Then, \emph DiRF predicts the best AP for the mobile client to connect with based on the mobile client's moving direction and AP's position. A ranking based algorithm is proposed to do AP selection, which can measure the similarity between the distribution of SNRs and the distribution of distances between the client and the associated AP. To decrease downlink packet retransmissions caused by AP switching, The central controller adds (explicit congestion notification) ECN markings to the downlink packets coming from the content server based on the mobile client's moving trajectory. As the mobile client moves toward the edge of associated AP's coverage, the content server will decrease downlink packets sent to the mobile client. There are four design challenges to achieve high network capacity in the indoor area with densely deployed APs.

First, we can achieve high throughput when the mobile client falls in the associated AP's FoV. However, each directional AP can just cover a small area. Therefore, it is important to achieve high throughput and seamless connectivity when the mobile client moves in the transition area of multiple directional APs. To address this challenge, we design a novel AP selection algorithm and a downlink packet scheduler to provide high throughput during AP switching.

Second, AP selection problem is investigated by many existing literatures~\cite{croitoru2015towards, song2017wi}. However, CSI based AP selection approaches cause frequent AP switching, which will degrade the network performance. To address this challenge, \emph DiRF selects the best AP based on the mobile client's moving direction, which will decrease the number of AP switching during movement.

Finally, the downlink packets which are not successfully forwarded by the current AP due to AP switching have to be retransmitted by the server after timeouts. To decrease the retransmissions caused by AP switching, the central controller adds ECN markings to the downlink packets based on the mobile client's moving trajectory.

\emph DiRF's contributions are three-fold as follows.
\begin{itemize}
\item We propose to use directional antenna mounted on the commercial AP to achieve high capacity communication when the client falls in the associated AP's FoV,  and robust connectivity when the client falls in the overlap of multiple APs' FoVs. 
\item We design an approach for central controller to add ECN markings to the downlink packets based on the mobile client's moving trajectory, which will decrease the downlink packets' retransmissions caused by AP switching. 
\item We implement \emph DiRF on a commercial-of-the-shelf (COTS) testbed with 6 directional antennas mounted on the 6 commercial APs. Our experiments show that \emph DiRF achieves a $3.16\times$ TCP throughput improvement, compared to the conventional scheme that uses one AP mounted with one omni-directional antenna. 
\end{itemize}

\section{Overview}
\label{section:overview}

\begin{figure}
\centering
    \includegraphics[width=0.8\linewidth]{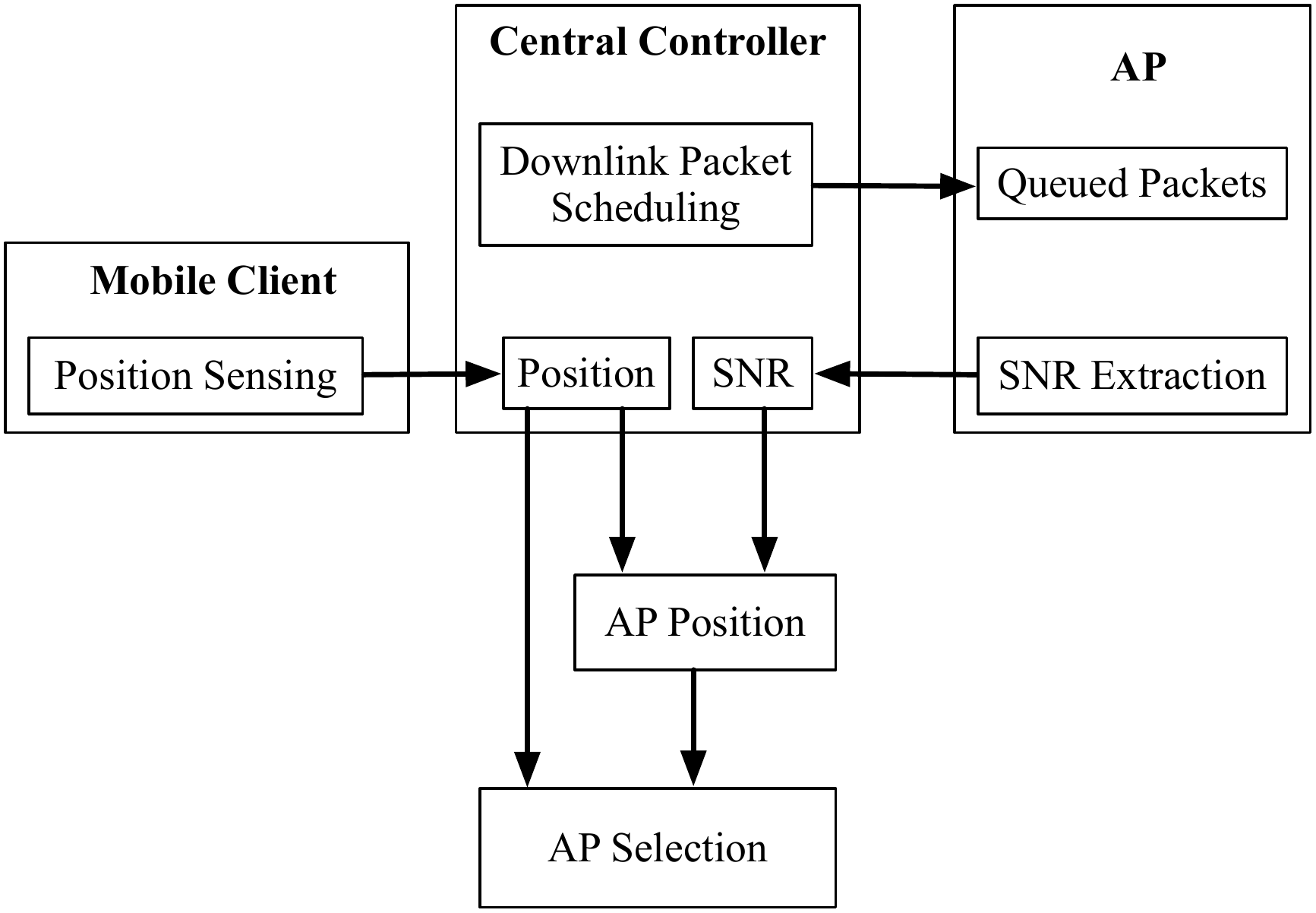}
    \caption{\emph DiRF Workflow: Position based AP selection is used during AP switching. And, ECN marking based downlink packet scheduler is used for decreasing the retransmissions caused by AP switching.}
    \label{fig:workflow}
\end{figure}

\emph DiRF adopts the workflow in Fig.~\ref{fig:workflow} to achieve high data rate transmission, when the mobile client moves in the indoor area covered by multiple deployed APs mounted with directional antennas. When the mobile client moves in one AP's coverage, the mobile client reports position information to the central controller through UDP packets. After the central controller estimates AP's position within the mobile client's coordinate system, \emph DiRF measures the similarity between the vector representing the AP's position and the vector representing the mobile client's moving direction in order to select the best AP for the mobile client to connect with. The central controller has known the relative distance among all APs. To estimate AP's position in the mobile client's coordinate system, \emph DiRF needs to estimate position of AP that the mobile client originally connects with. Therefore, when the mobile client moves in the coverage of one AP that the mobile client originally connects with, the central controller extracts SNR from the associated AP. And, the central controller has the mobile client's position information. Then, the central controller can estimate the position of AP that the mobile client originally connects with based on the distribution of collected SNR measurements and the distribution of distances between the mobile client and the associated AP.

During AP switching, the downlink packets which are not successfully forwarded by the current AP have to be retransmitted by the server after timeouts, which will degrade the network performance. To decrease the retransmissions caused by AP switching, the central controller adds ECN markings to the downlink packets based on the mobile client's moving trajectory. The ECN markings can notify the server to decrease downlink packets transmitted to the mobile client, when the mobile client moves toward the edge of the associated AP's coverage. Therefore, the retransmissions of downlink packets after AP switching are decreased.

\section{System Design}
\label{section:design}

In this section, we discuss \emph DiRF's design, which consists of two components. First, we design an algorithm to select best AP for the mobile client based on the mobile client's moving direction. Second, we design a downlink packet scheduler in order to decrease the retransmissions caused by AP switching.

\subsection{AP Selection}
\label{subsection:aps}

The existing work selects the best AP based on the link state or AP's state, which is measured by metrics such as SNR~\cite{imran2016service}, ESNR~\cite{song2017wi} and load balancing of AP~\cite{bhartiaclientmarshal}. However, these metrics reflect the current link state or AP's state, and selecting AP based on the current link state or AP's state causes frequent AP switching in the near future without considering the mobile client's moving direction. And, the frequent AP switching causes latency accumulation, which degrades the network throughput. Therefore, \emph DiRF selects the best AP based on the mobile client's moving direction.

\begin{figure*}
\centering
\begin{subfigure}{0.32\textwidth}
  \centering
  \includegraphics[width=0.8\linewidth, height=0.6\linewidth]{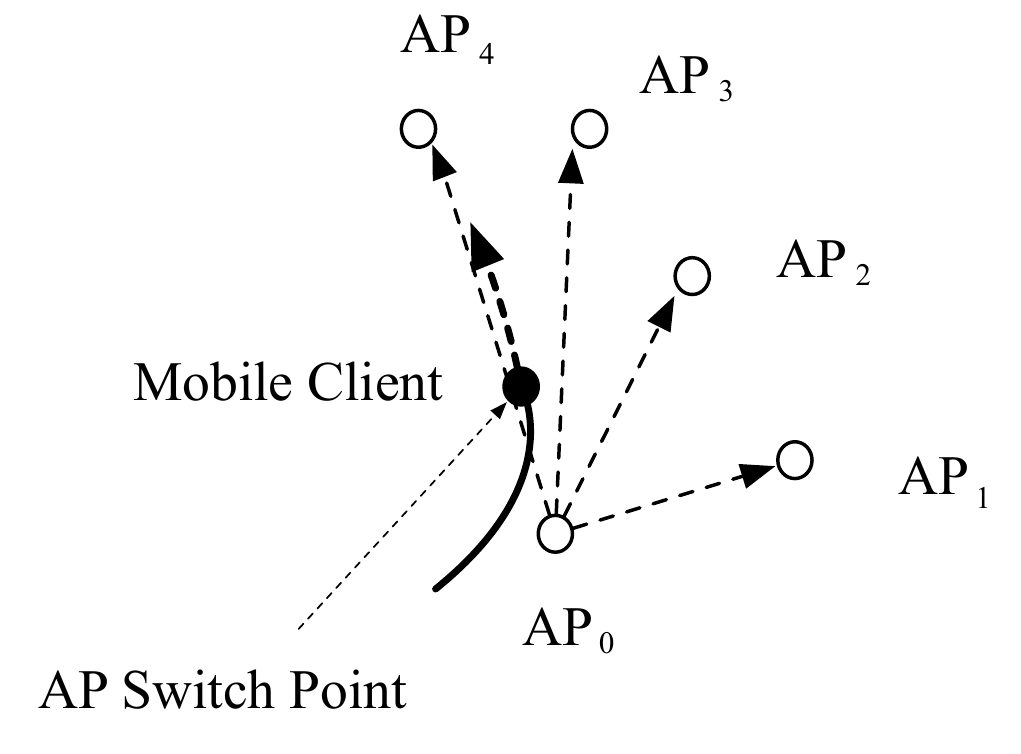}
  \caption{AP Selection.}
  \label{fig:ap_selection}
\end{subfigure}%
\begin{subfigure}{0.32\textwidth}
  \centering
  \includegraphics[width=0.8\linewidth, height=0.6\linewidth]{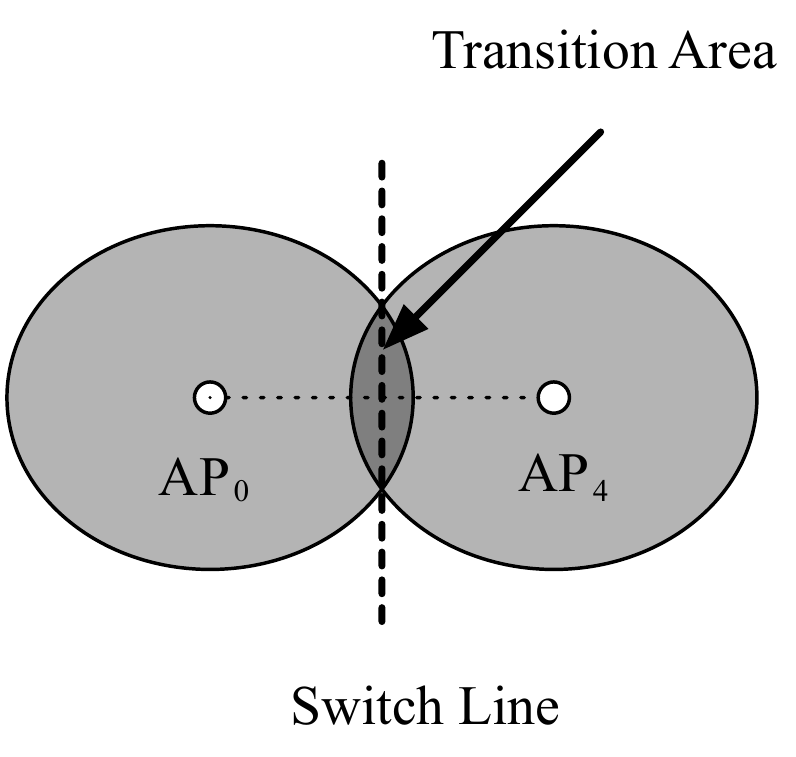}
  \caption{AP Switching}
  \label{fig:switch_line}
\end{subfigure}%
\begin{subfigure}{0.32\textwidth}
    \centering
    \includegraphics[width=0.8\linewidth, height=0.6\linewidth]{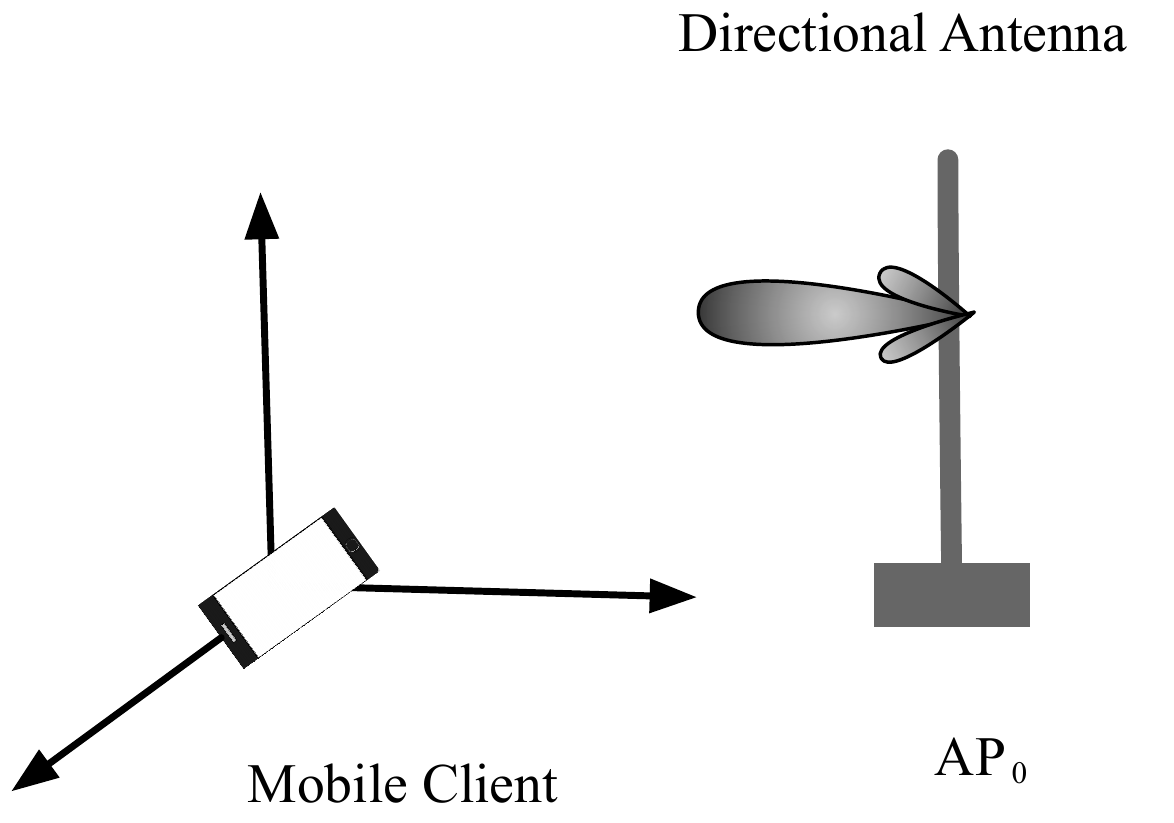}
    \caption{Position Estimation.}
    \label{fig:position_estimation}
\end{subfigure}
\caption{AP Selection: (\subref{fig:ap_selection}) AP selection is based on the mobile client's moving direction. (\subref{fig:switch_line}) AP switching at the switch line, which is orthogonal to the line between two APs in the transition area of multiple adjacent APs. (\subref{fig:position_estimation}) $AP_{0}$'s position is estimated within the mobile client's coordinate system.}
\label{fig:ap_selection_line}
\end{figure*}

The position information is denoted as $\left [ x, y, z \right ]$, where x, y and z represent 3D location of the mobile client. So, $\mathbf{p_{m}}=\left [ x_{m}, y_{m}, z_{m} \right ]$ represents the mobile client's position and $\mathbf{p_{a}}=\left [ x_{a}, y_{a}, z_{a}\right ]$ represents AP's position. To predict the mobile client's moving direction in comparison to AP's position, \emph DiRF measures the similarity between the vector representing the mobile client's moving direction and the vector representing AP's position. The mobile client originally connects with one AP ($AP_{0}$) with position $\mathbf{p_{a_{0}}}=\left [ x_{a_{0}}, y_{a_{0}}, z_{a_{0}}\right ]$. And, the position of $AP_{i}, i\neq0$ is represented by $\mathbf{p_{a_{i}}}=\left [ x_{a_{i}}, y_{a_{i}}, z_{a_{i}}\right ]$. The mobile client's moving direction is represented by the variation of mobile client's position during the short time period $\left [ t, {t}' \right ]$, which is defined as follows:
\begin{equation}
\mathbf{p_{m_{t{t}'}}}=\left [ x_{m_{{t}'}}, y_{m_{{t}'}}, z_{m_{{t}'}} \right ]
-\left [ x_{m_{t}}, y_{m_{t}}, z_{m_{t}} \right ].
\end{equation}
Therefore, the best AP can be selected by solving the following problem:
\begin{equation}
\label{equation:apindex}
a^{*}=\underset{i\in A}{argmin} \left \| \mathbf{p_{m_{t{t}'}}}-\mathbf{p_{a_{i}}} \right \|_{2}^{2},
\end{equation}
where $A$ is a set containing all APs' indexes, $a^{*}$ is the best AP's index. Fig.~\ref{fig:ap_selection_line}~(\subref{fig:ap_selection}) shows the AP selection based on the mobile client's moving direction. And, the mobile client currently connects with $AP_{0}$. At the AP switching point, the central controller has to make a decision on which AP the mobile client should connect with based on the similarity between the vector representing the mobile client's moving direction and the vector representing AP's position. To reduce the latency accumulation, the mobile client connects with $AP_{4}$ along its future moving direction rather than the other APs ($AP_{1}, AP_{2}, AP_{3}$) with slightly higher current SNR. Fig.~\ref{fig:ap_selection_line}~(\subref{fig:switch_line}) shows the switch line indicating that the mobile client has to switch AP and the transition area of multiple adjacent APs.

In the above discussion, we assume that the mobile client's position and AP's position are in the same coordinate system. In practice, the central controller knows each AP's position. And, the mobile client's position is measured by the embedded sensors of the mobile client. However, the mobile client's position and AP's position are not measured in the same coordinate system. Therefore, the central controller estimates AP's position within the mobile client's coordinate system in Fig.~\ref{fig:ap_selection_line}~(\subref{fig:position_estimation}).  When the mobile client moves around in the coverage of $AP_{0}$, the mobile client uploads its position information to the central controller through UDP packets. And, the central controller gets the SNR value from AP forwarding the mobile client's packets.

To estimate the position of $AP_{0}$, the central controller collects a set of SNR samples $R$ and a set of corresponding 3D locations $Q$ of the mobile client. We know that the higher SNR value indicates the shorter distance between the mobile client and the associated AP. However, SNR value cannot reflect the distance between transmitter and receiver accurately~\cite{han2013localization} due to the frequency selective channel. Therefore, we leverage the relationship between the distribution of SNR values and the distribution of distances between the mobile client and the associated AP to estimate the position of $AP_{0}$. $P\left ( \left \langle r_{j},  r_{k} \right \rangle \right )=\frac{ r_{j}}{ r_{j}+ r_{k}}$ is the probability of j-th SNR sample $r_{j}$ being ranked higher than k-th SNR sample $r_{k}$. We use $d_{j}$ and $d_{k}$ to represent the j-th distance sample and k-th distance sample calculated by $d_{j}=\left \|  \mathbf{p_{a_{0}}} -\mathbf{p_{m_{j}}} \right \|_{2}^{2}$ and $d_{k}=\left \|  \mathbf{p_{a_{0}}} -\mathbf{p_{m_{k}}} \right \|_{2}^{2}$ respectively. $\mathbf{p_{m_{j}}}$ and $\mathbf{p_{m_{k}}}$ are j-th 3D location and k-th 3D location of the mobile client. Then, $P\left ( \left \langle d_{j},  d_{k} \right \rangle \right )=\frac{d_{k}}{d_{k}+d_{j}}$ is the probability of j-th distance sample being ranked higher than k-th distance sample. Then, we define our loss function based on the cross-entropy loss, which can measure the similarity between the distribution of SNR samples and the distribution of the distances between the mobile client and the associated AP. Therefore, we estimate the position of $AP_{0}$ by solving the following optimization problem:
\begin{eqnarray}
\label{equation:ap:pose}
\mathbf{p_{a_{0}}^{*}}=\underset{\mathbf{p_{a_{0}}}\in R^{3}}{argmin}-\frac{1}{s}\sum_{i=1}^{s}P\left ( \left \langle r_{j},  r_{k} \right \rangle_{i} \right )logP\left ( \left \langle d_{j},  d_{k} \right \rangle_{i} \right )\\
\nonumber
+\left ( 1- P\left ( \left \langle r_{j},  r_{k} \right \rangle_{i} \right )\right )log\left ( 1-P\left ( \left \langle d_{j},  d_{k} \right \rangle_{i} \right ) \right ),
\end{eqnarray}
where $s$ is the size of SNR samples or mobile client's position samples, $\mathbf{p_{a_{0}}^{*}}$ is the estimated position of $AP_{0}$ within the mobile client's coordinate system. And, Eq.~(\ref{equation:ap:pose}) is a convex optimization, which can be solved by the gradient descent method \cite{boyd2004convex}.

Algorithm~\ref{Alg:ap:selection} summarizes the procedure of selecting best AP based on the mobile client's moving direction. When the mobile client moves around in the coverage of one AP, the mobile client records the position and uploads to the central controller. And, the central controller gets the SNR value from AP forwarding the mobile client's packets. When the mobile client arrives at the switch line in the step 5, the mobile client has to switch to the another AP. And, the central controller calculates position of $AP_{0}$ in the step 6. Then, central controller can compute all APs' positions in the step 7 based on the known relative distance among all APs with $G(\cdot )$, where $G(\cdot )$ is a function describing the relative distance among all APs. To make a decision on which AP the mobile client has to connect with, the central controller  estimates the mobile client's moving direction in comparison to all APs' positions from the step 8 to 13. At last, the central controller selects $AP_{a^{*}}$ as the best AP to switch to in the step 14.
The latency of AP selection is important for AP handoff. And, the computational complexity of Algorithm \ref{Alg:ap:selection} during AP switching (from step 8 to step 14) is $O\left ( \left | A \right | \right )$, which cannot cause much latency during AP switching.

\begin{algorithm}
\caption{AP Selection}\label{Alg:ap:selection}
\begin{algorithmic}[1]
\Procedure{APSelection}{$Q$}
\State $dMin\gets0$ \Comment{initialization}
\While{($Q$)}
\State $d\gets Q$ \Comment{read the current position}
\If{$f(d)=0$} \Comment{$f(\cdot )=0$ is the switch line}
\State  $\mathbf{p_{a_{0}}} \gets \mathbf{p_{a_{0}}^{*}}$ \Comment{find the position of $AP_{0}$ with Eq.~(\ref{equation:ap:pose})}
\State $\mathbf{p_{a_{i}}} \gets G\left (  \mathbf{p_{a_{0}}}\right ), i\in A$ \Comment{Calculating all AP's positions with $G(\cdot )$}
\For{$i\in A$}  \Comment{Eq.~(\ref{equation:apindex})}
\If{$dMin \geq \left \| \mathbf{p_{m_{t{t}'}}}-\mathbf{p_{a_{i}}} \right \|_{2}^{2}$}
\State $a^{*} \gets i$
\State $dMin = \left \| \mathbf{p_{m_{t{t}'}}}-\mathbf{p_{a_{i}}} \right \|_{2}^{2}$
\EndIf
\EndFor
\State \textbf{return} $a^{*}$ \Comment{return the index of best AP}
\Else
\State Continue \Comment{AP switching is not triggered}
\EndIf
\EndWhile
\EndProcedure
\end{algorithmic}
\end{algorithm}

\subsection{Downlink Packet Scheduler}
\label{subsection:scheduler}

The mobile client's traffic, especially the smartphone traffic, represents a large amount of Internet traffic. And, the mobile client's traffic contains downlink traffic and uplink traffic such as online video streaming~\cite{liu2008survey} and web browsing~\cite{lieberman1995letizia}. The ratio of downlink traffic and uplink traffic caused by the mobile client is very unbalanced. The authors in~\cite{falaki2010first} show that the ratio of downlink traffic and uplink traffic for smartphones is 6:1. This high unbalance indicates the strong bias towards downloads on the mobile client. Therefore, we focus on the downlink packets in this paper.

When the content server transmits packets to the mobile client, the packets are cached in AP's buffer until they are forwarded by the current AP. And, when the mobile client moves toward the edge of the current AP's coverage, the downlink packets are prone to loss due to AP switching and wireless multipath fading. Then, the server has to retransmit these packets after timeouts. To decrease these retransmissions, WiFi Goes to Town \cite{song2017wi} multicasts the downlink packets to all the cooperating APs and acknowledges the downlink packets with block acknowledgement. However, the same downlink packet has to be stored in each AP's buffer, which increases the overhead and the latency due to the communication among all the coordinating APs. Instead of duplicating the downlink packets to each AP, the central controller can forward the downlink packets with considering the mobile client's future mobility such that the number of downlink packets cached in the current AP's buffer is minimized before AP switching. Therefore, we can regard the central controller as a router to manage all the downlink packets transmitted by the content server and forward all the downlink packets to the best AP.

\emph DiRF uses explicit congestion notification (ECN) to tell the content server that the mobile client is going to be near the edge of current AP's FoV. Then, the content server will decrease the rate of sending downlink packets to the current AP. And, the number of downlink packets cached in the current AP's buffer becomes smaller, when the mobile client moves toward the edge of current AP's FoV. Therefore, the central controller has to add the ECN markings to the downlink packets such that there are minimal number of downlink packets cached in the current AP's buffer before AP switching.

To have minimized number of downlink packets cached in the current AP's buffer before AP switching, we estimate the rate of ECN markings that the central controller needs to add to the donwlink packets based on the mobile client's moving trajectory. Then, we use a simple throughput model $B=\sqrt{\frac{2}{p+\alpha}}\frac{MSS}{RTT}$ \cite{croitoru2015towards} to estimate the throughput, where $B$ is the throughput, $\alpha$ is the rate of adding ECN markings and $p$ is the packet's loss rate. The maximum segment size (MSS) is a parameter in the TCP header. The time interval that the mobile client has to switch to another AP is $\Delta t = \frac{d}{v}$, where $d$ is the distance between the mobile client and the switch line, and $v$ is the mobile client's speed which can be calculated through the mobile client's sensing data. Then, we can calculate the rate of adding ECN markings as follows:
\begin{equation}
\label{Eq:ecn_marking}
B\Delta t = b \Rightarrow \alpha=2(\frac{MSS\Delta t}{bRTT})^{2}-p, 
\end{equation}
where $b$ is the size of AP's buffer. To estimate the value of $b$, we know that nearly all wireless APs can buffer the data to provide 802.11a/g capacity, which is 25Mbps for the mobile client to download data from the server with the RTT at least 100ms~\cite{croitoru2015towards}. Therefore, the smallest size of the buffer is 2.5Mbits. Note that we give the maximum value of $\alpha$ in Eq.~(\ref{Eq:ecn_marking}). And, the marking rate we use in our experiment is lower than the marking rate in Eq.~(\ref{Eq:ecn_marking}).

\section{Implementation}
\label{section:implementation}

\begin{figure}
    \centering
    \includegraphics[width=0.8\linewidth]{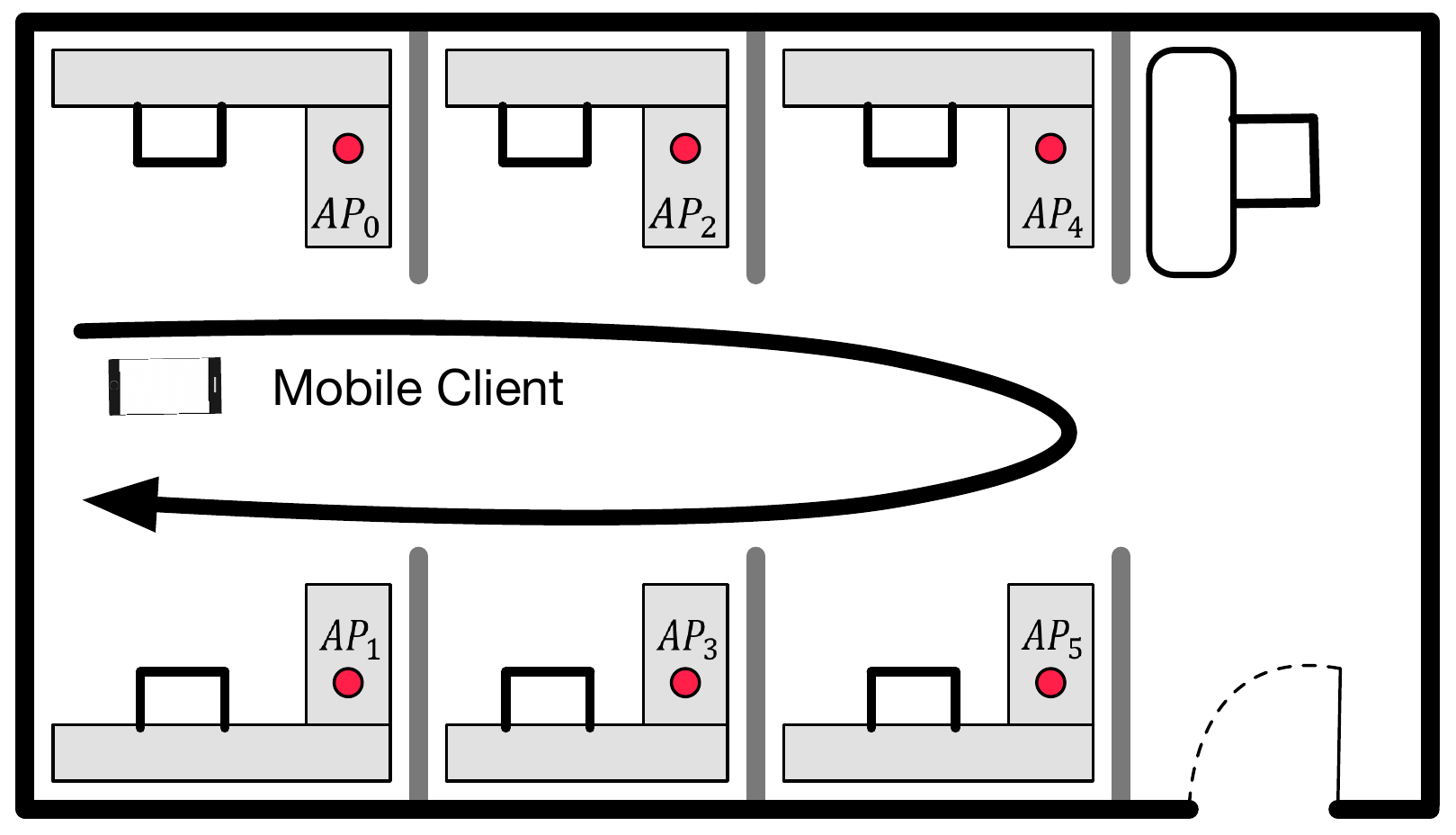}
    \caption{Experiement Setup: We deploy APs in the room on the first floor of an office building. And, the mobile client moves in the following way: $AP_{0}\rightarrow AP_{2}\rightarrow AP_{4}\rightarrow AP_{5}\rightarrow AP_{3}\rightarrow AP_{1}$.}
    \label{fig:floor_plan}
\end{figure}

We prototype \emph DiRF based on the commercial APs mounted with the directional antennas. The testbed has six APs mounted with six directional antennas, which are connected with the central controller through the Ethernet backhaul. Also, the central controller is connected with the content server. We deploy the testbed in the laboratory. Each AP mounted with the directional antenna can serve the mobile client in an area of the laboratory as shown in Fig.~\ref{fig:floor_plan}. 

\textbf{Central Controller.} We use HP EliteBook 8470p laptop as the central controller, which is equipped with 3rd Generation Intel Core i7 CPU, 256 GB SED Solid State Drive (SSD) and 16G DDR3 RAM. We can regard the central controller as a router to schedule the packets and make a decision on the best AP selection. Also, The central controller is installed with the Click modular router~\cite{kohler2000click} on Ubuntu Linux v16.04 LTS. We write a click configuration for central controller's control logic in order to add ECN markings to the downlink packets and forward the downlink packets to the best AP. Therefore, the central controller receives the downlink packets from the content server, and schedules the downlink packets to the best AP based on the AP selection (\S \ref{subsection:aps}) and downlink packet scheduler (\S \ref{subsection:scheduler}).

\textbf{Directional AP.} We use TP-Link N750 AP~\cite{ap} mounted with TL-ANT2409A 2.4GHz 9dBi Directional Antenna~\cite{antenna}. We detach AP's omni-directional antennas and use a Mini-Circuits ZA3PD-4-S+ RF splitter-combiner to connect AP with the directional antenna. All the APs are connected with the TP-Link Ethernet Switch, which is then connected with the central controller to formulate a LAN.

\textbf{Mobile Client.} Our mobile client is Motorola Moto X (2nd Gen) equipped with Android version 5.1, Quad-core 2.5GHz Krait 400 and 16GB storage. We design an Android application to collect the sensing data and forward the sensing data to the central controller with UDP packets. 

\section{Evaluation}
\label{section:evaluation}

We first evaluate the end-to-end performance of \emph DiRF. Then, we present micro-benchmark experiments to see which factors impact \emph DiRF's performance. After that, we conduct two real-world case studies to show \emph DiRF's performance of handling web downloads and web browsing. 

\subsection{Methodology}

We deploy six commercial access points mounted with directional antennas on the ceiling of room. And, the mobile client moves in the APs' coverages at the moderate walking speed (around 3.13mph). The mobile client first connects with $AP_{0}$. Then, the mobile client moves toward APs in the following way as shown in Fig.~\ref{fig:floor_plan}: $AP_{0}\rightarrow AP_{2}\rightarrow AP_{4}\rightarrow AP_{5}\rightarrow AP_{3}\rightarrow AP_{1}$. We use iperf3~\cite{iperf3} to measure \emph DiRF's performance for micro-benchmark experiments and use tcpdump to log packet flows for data analysis in two real-world case studies. And, our experiments follow 802.11 standard working at 5GHz on 20MHz bandwidth as default.

\textbf{Comparison scheme.} We use one AP mounted with an omni-directional antenna to cover the whole place of a room for comparison. Currently, this comparison scheme is adopted widely in the home and office. We term this scheme \emph OmRF, using it as a comparison for our evaluation.

\subsection{End-to-end Performance}

We now evaluate the end-to-end performance of \emph DiRF delivering TCP and UDP data flows, when there is one mobile client moving around.

\begin{figure*}
\begin{minipage}[t]{.24\textwidth} 
    \includegraphics[width=\textwidth]{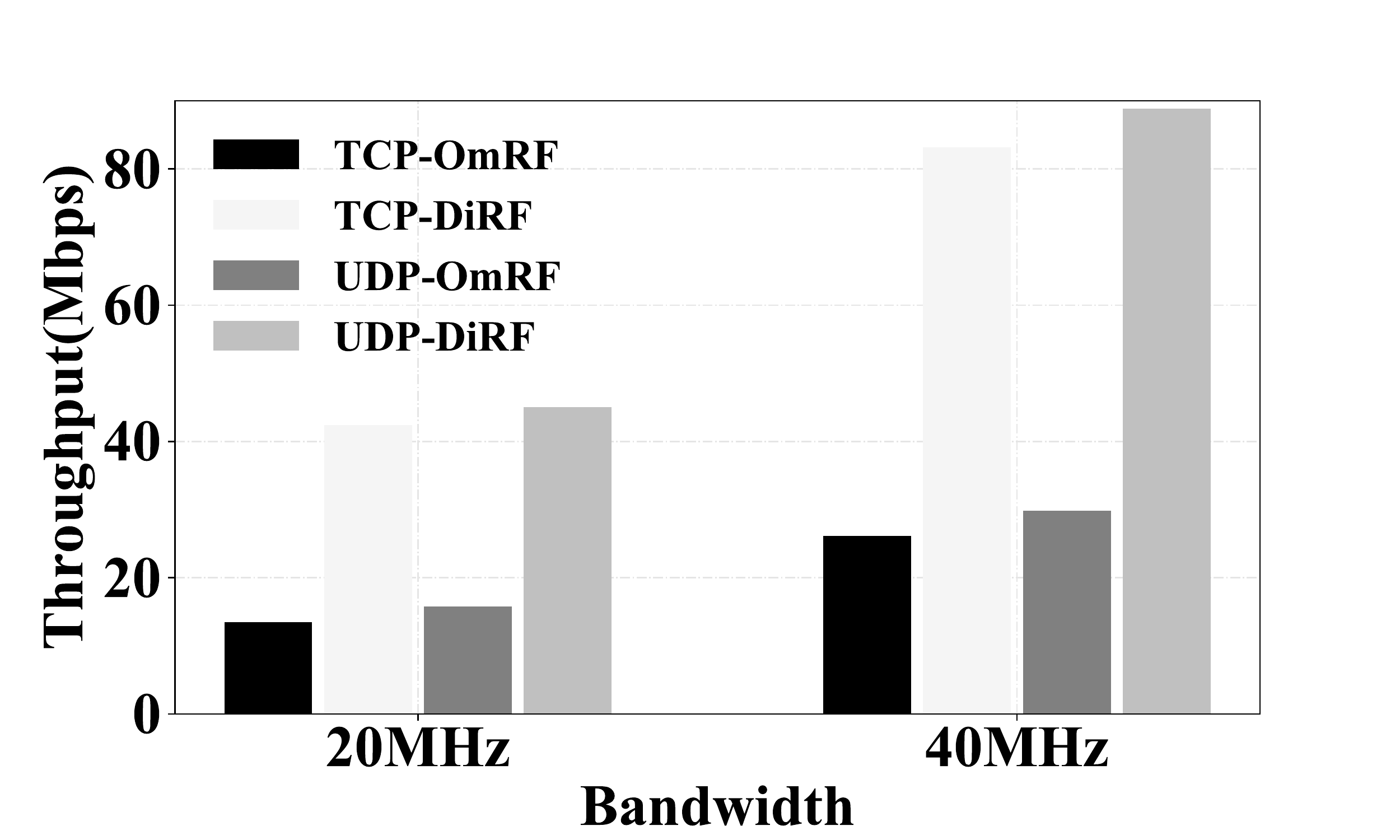}
   \caption{TCP and UDP throughput over different bandwidth.}
    \label{fig:end_to_end}
\end{minipage}%
\hfill 
\begin{minipage}[t]{.24\textwidth}
    \includegraphics[width=\linewidth]{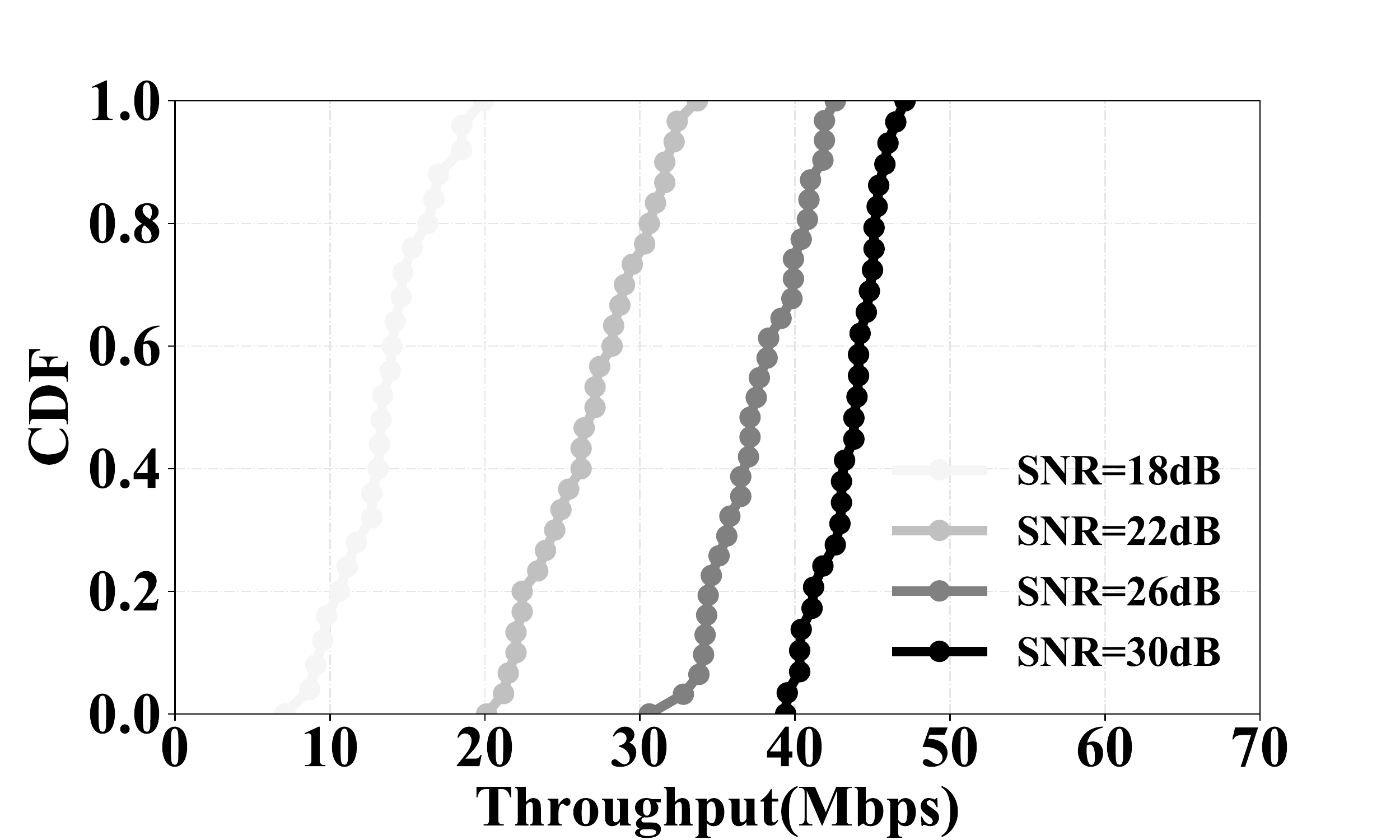}
    \caption{TCP throughput when varying SNR values.}
    \label{fig:throughput_snr}
\end{minipage}%
\hfill
\begin{minipage}[t]{.24\textwidth} 
    \includegraphics[width=\textwidth]{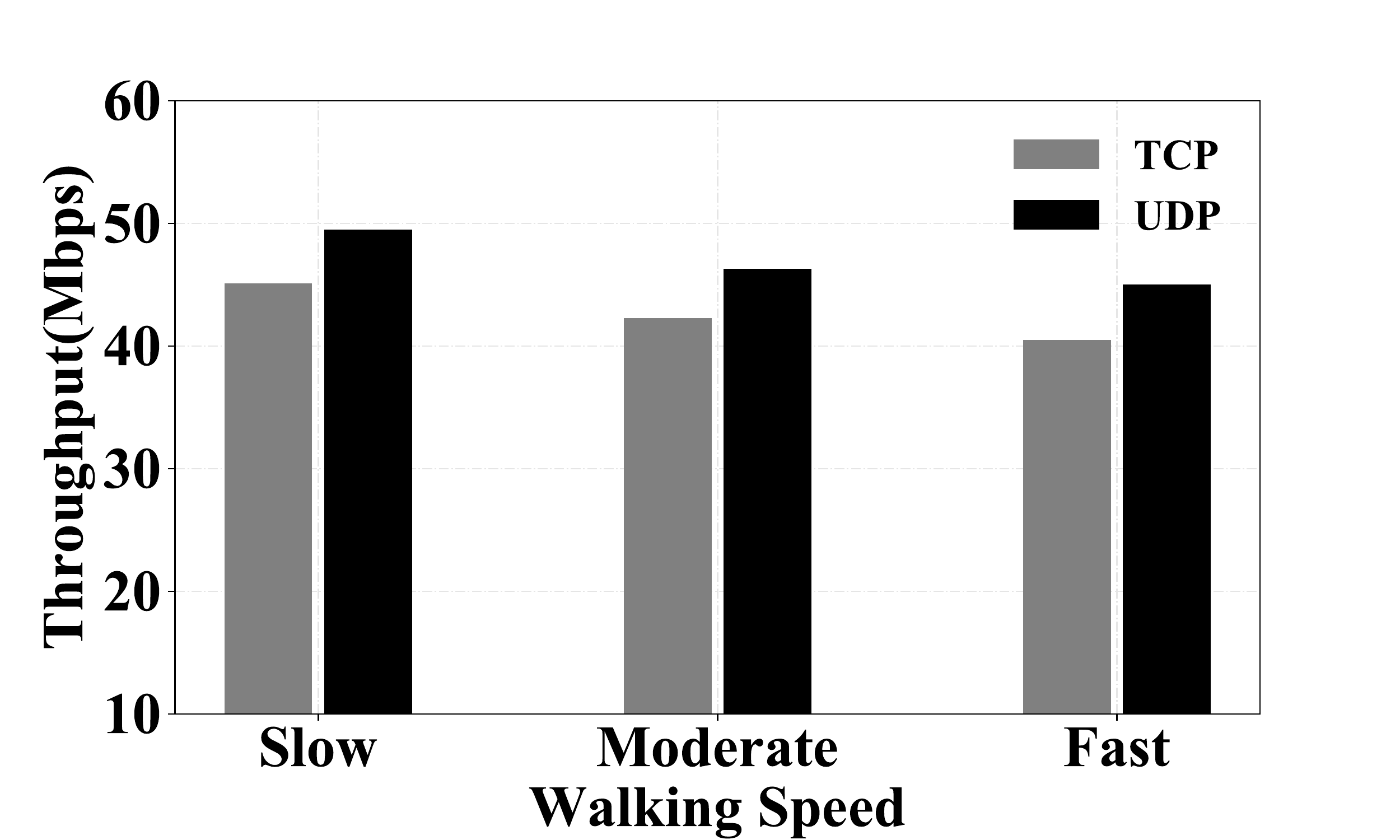}
    \caption{Average throughput over different walking speeds.}
    \label{fig:speed_throughput}
\end{minipage}%
\hfill 
\begin{minipage}[t]{.24\textwidth}
    \includegraphics[width=\textwidth]{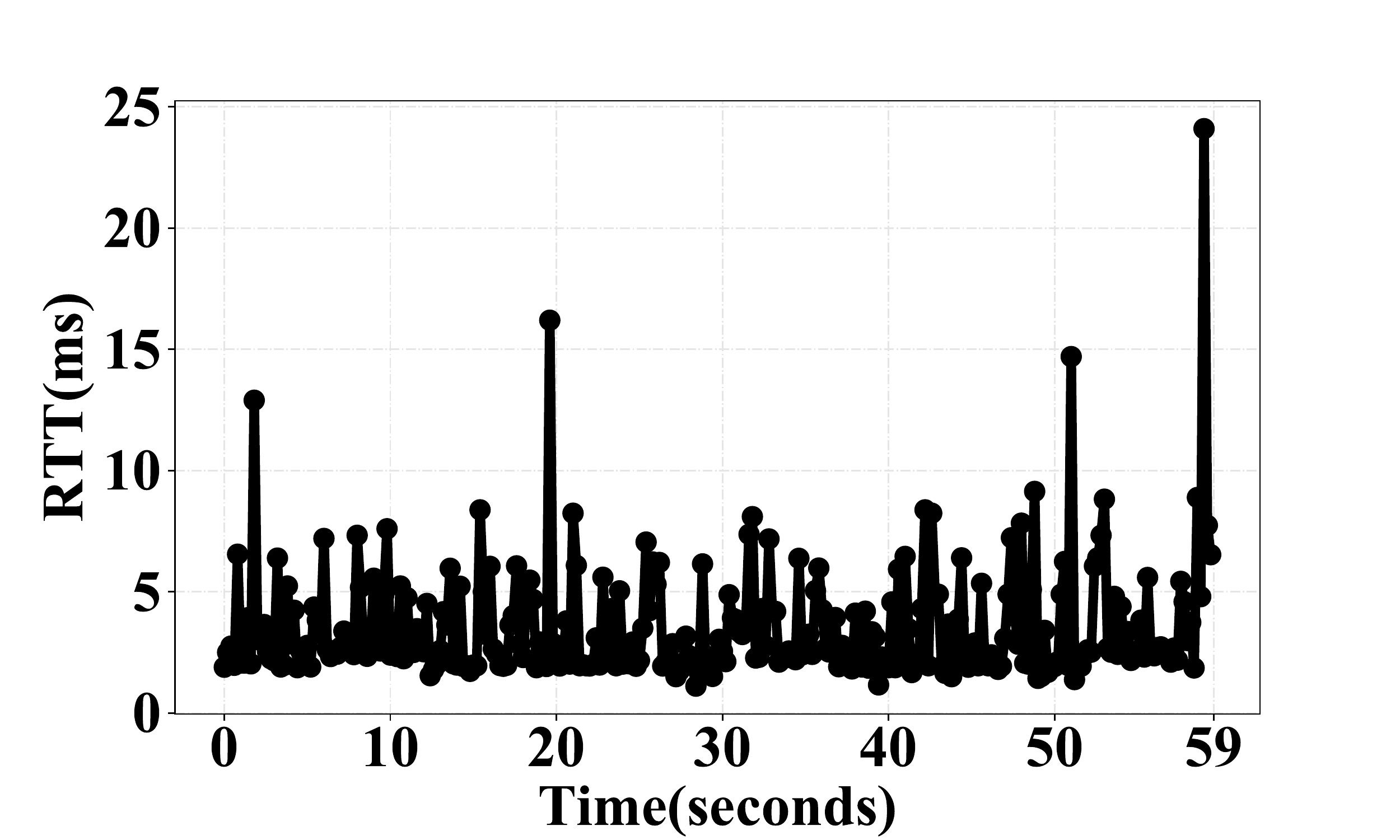}
    \caption{Packet latency during AP switching.}
    \label{fig:ap_latency}
\end{minipage}

\end{figure*}

In Fig.~\ref{fig:end_to_end}, we plot the average TCP and UDP throughput of \emph DiRF and \emph OmRF on both 20MHz and 40MHz bandwidth. For 20MHz bandwidth, the average TCP throughput of \emph DiRF is $3.16\times$ the average TCP throughput of \emph OmRF. And, the average UDP throughput of \emph DiRF is $2.85\times$ the average UDP throughput of \emph OmRF. For 40MHz bandwidth, the average TCP throughput of \emph DiRF is $3.18\times$ the average TCP throughput of \emph OmRF. And, the average UDP throughput of \emph DiRF is $2.76\times$ the average UDP throughput of \emph OmRF. This is because \emph DiRF can improve the spatial reuse. Also, we can see that the average throughput increases as the bandwidth increases. The average TCP throughput of \emph DiRF on 40MHz is $1.96\times$ the average TCP throughput of \emph DiRF on 20MHz. And, the average UDP throughput of \emph DiRF on 40MHz is $1.97\times$ the average UDP throughput of \emph DiRF on 20MHz. Therefore, \emph DiRF can achieve high capacity communication in comparison to \emph OmRF.

\subsection{Micro Benchmarks}
We now present micro benchmarks to understand the impact of factors on our system's performance. We first test the TCP throughput against SNRs. We also examine TCP and UDP performance against different walking speeds of the mobile client. Then, we evaluate the error of estimating $AP_{0}$'s position, the accuracy of AP selection and the AP switching latency. After that, we test the TCP and UDP performance varying the number of mobile clients and AP density.

\textbf{Impact of SNR.} With directional antenna mounted on AP, the high throughput can be achieved easily when the receiver falls in AP's FoV. As shown in Fig.~\ref{fig:throughput_snr}, we plot the CDF of throughput when varying SNR values. Our experiment follows 802.11 standard working at 5GHz on 20MHz bandwidth. When the SNR value is 18dB, the median throughput is 15Mbps. When the SNR value is 30dB, the median throughput is 45Mbps. The throughput increases as the SNR increases. Therefore, high capacity communication can be achieved with high gain directional antenna.

\textbf{Impact of the mobile client's walking speed.} We examine the average TCP and UDP throughput, when one mobile client moves in APs' coverages at the different walking speeds. As shown in Fig.~\ref{fig:speed_throughput}, when the mobile client moves at a moderate speed (around 3.13mph), the average TCP and UDP throughput can achieve 42.3Mbps and 46.3Mbps respectively. When the mobile client moves at a slow speed (around 1.79mph), the average TCP and UDP throughput can achieve 45.1Mbps and 49.5Mbps respectively. When the mobile client moves at a fast speed (around 4.47mph), the average TCP and UDP throughput can achieve 40.5Mbps and 45Mbps respectively. As the walking speed becomes faster, the throughput becomes slightly lower. This is because the higher mobility causes more packet loss. However, the high throughput can be achieved with different walking speeds due to densely deployed APs mounted with high-gain directional antennas.

\begin{figure}
\centering
\begin{subfigure}{0.25\textwidth}
  \centering
  \includegraphics[width=1.0\linewidth, height=0.8\linewidth]{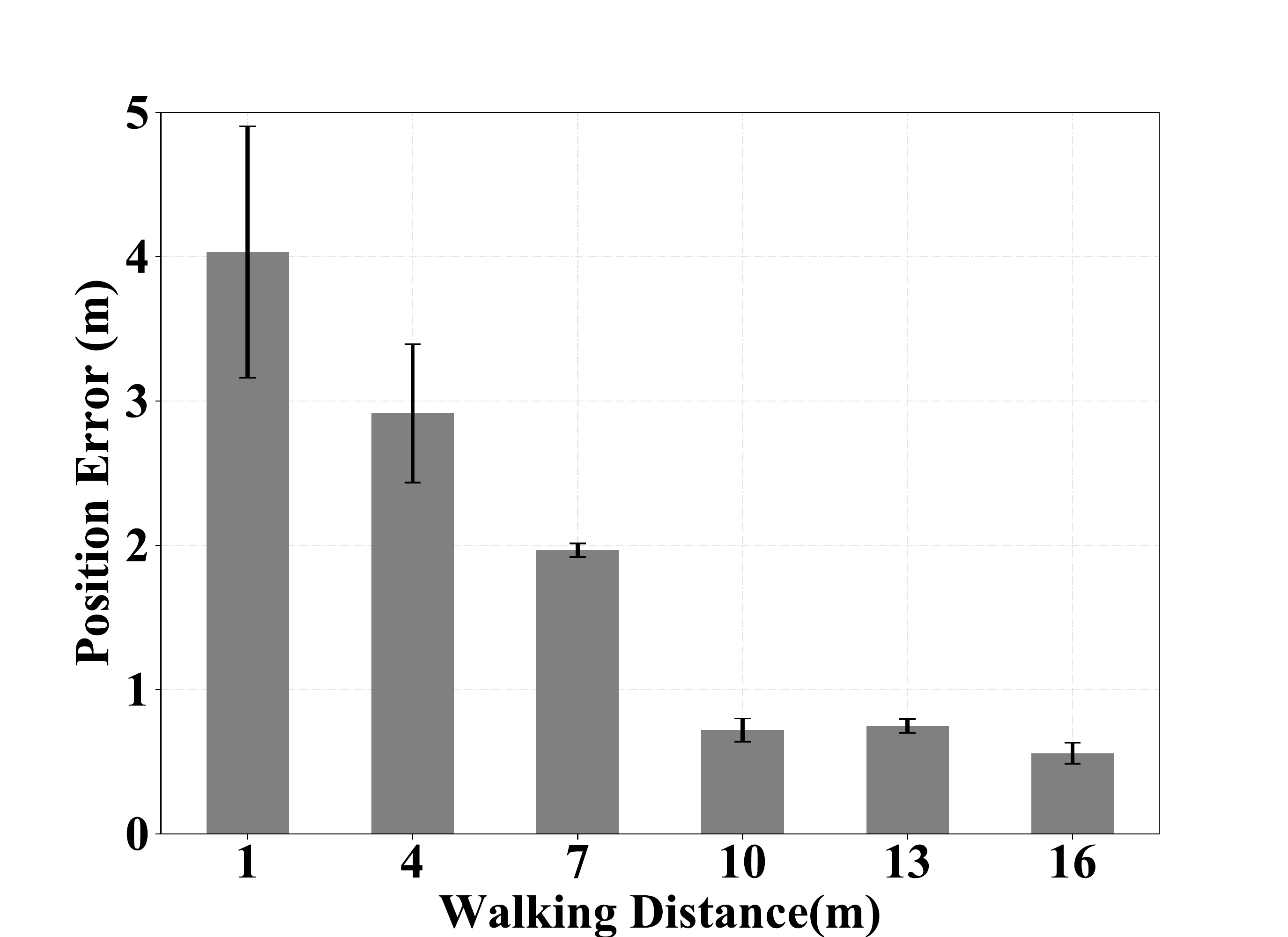}
  \caption{Error of the Position Estimation.}
  \label{fig:position_error}
\end{subfigure}%
\begin{subfigure}{0.25\textwidth}
  \centering
  \includegraphics[width=1.0\linewidth, height=0.8\linewidth]{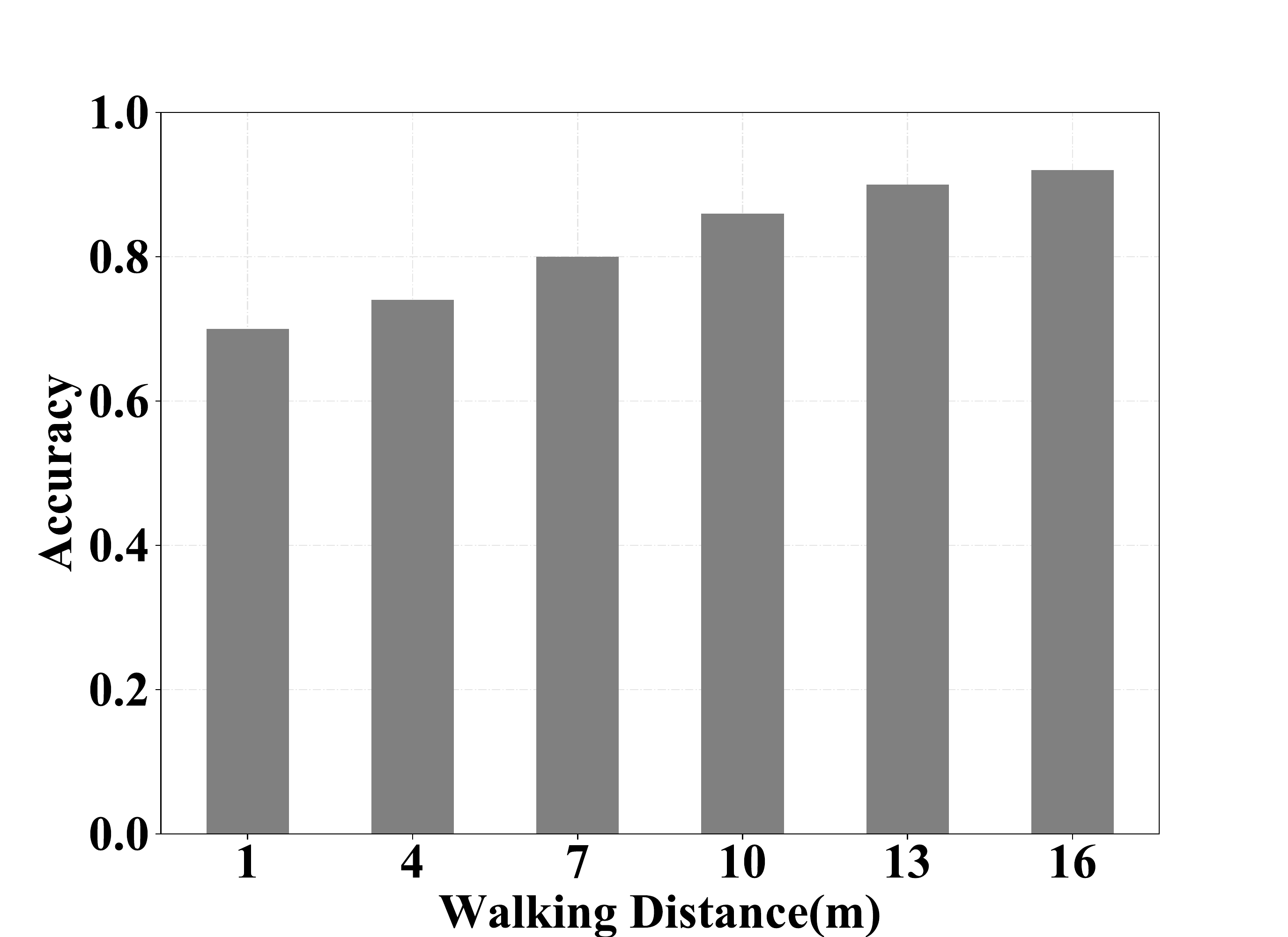}
  \caption{Accuracy of the AP Selection.}
  \label{fig:ap_selection_accuracy}
\end{subfigure}
\caption{Error of AP position estimation and accuracy of AP selection against the mobile client's walking distance.}
\label{fig:ap_sensing}
\end{figure}

\textbf{AP selection.} Our AP selection algorithm is based on the estimation of $AP_{0}$'s position within the mobile client's coordinate system and the mobile client's position. And, the central controller knows the relative distance among all APs. Then, we can predict the mobile client's moving direction in comparison to AP's position. Fig.~\ref{fig:ap_sensing}~(\subref{fig:position_error}) shows the error of $AP_{0}$'s position estimation, when the mobile client walks in $AP_{0}$'s FoV. We can see that the estimation error becomes smaller as the mobile client walks longer in AP's FoV. And, when the mobile client walks longer than 15m, the position estimation error is smaller than 0.5m. This is because the position estimation is based on the distribution of collected SNR values and the distribution of the measured distances between the mobile client and the associated AP. Fig.~\ref{fig:ap_sensing}~(\subref{fig:ap_selection_accuracy}) shows the accuracy of AP selection, when the mobile client moves around in the room. The mobile client first moves in $AP_{0}$'s coverage. When the mobile client moves to multiple APs' overlaps, the central controller has to make a decision on which AP the mobile client has to connect with based on Algorithm~\ref{Alg:ap:selection}. As shown in Fig.~\ref{fig:ap_sensing}~(\subref{fig:ap_selection_accuracy}), the accuracy of AP selection is increased as the walking distance in $AP_{0}$'s coverage increases. And, the accuracy of AP selection achieves 0.92, when the mobile client walks longer than 15m in $AP_{0}$'s coverage. This is because we can collect more SNR values and more mobile client's positions to predict the mobile client's moving direction accurately. 

\textbf{AP switching latency.} To verify seamless connectivity during AP switching, we enforce the mobile client to switch between two APs. As shown in Fig.~\ref{fig:ap_latency}, we plot the round trip time latency between the mobile client and the content server. When measurement occurs during a beacon interval, the latency is less than 10ms. We can see that the packet latency is larger than 10ms at four AP switching points due to the overhead.

\begin{figure}
\centering
\begin{subfigure}{0.25\textwidth}
  \centering
  \includegraphics[width=1.0\linewidth, height=0.8\linewidth]{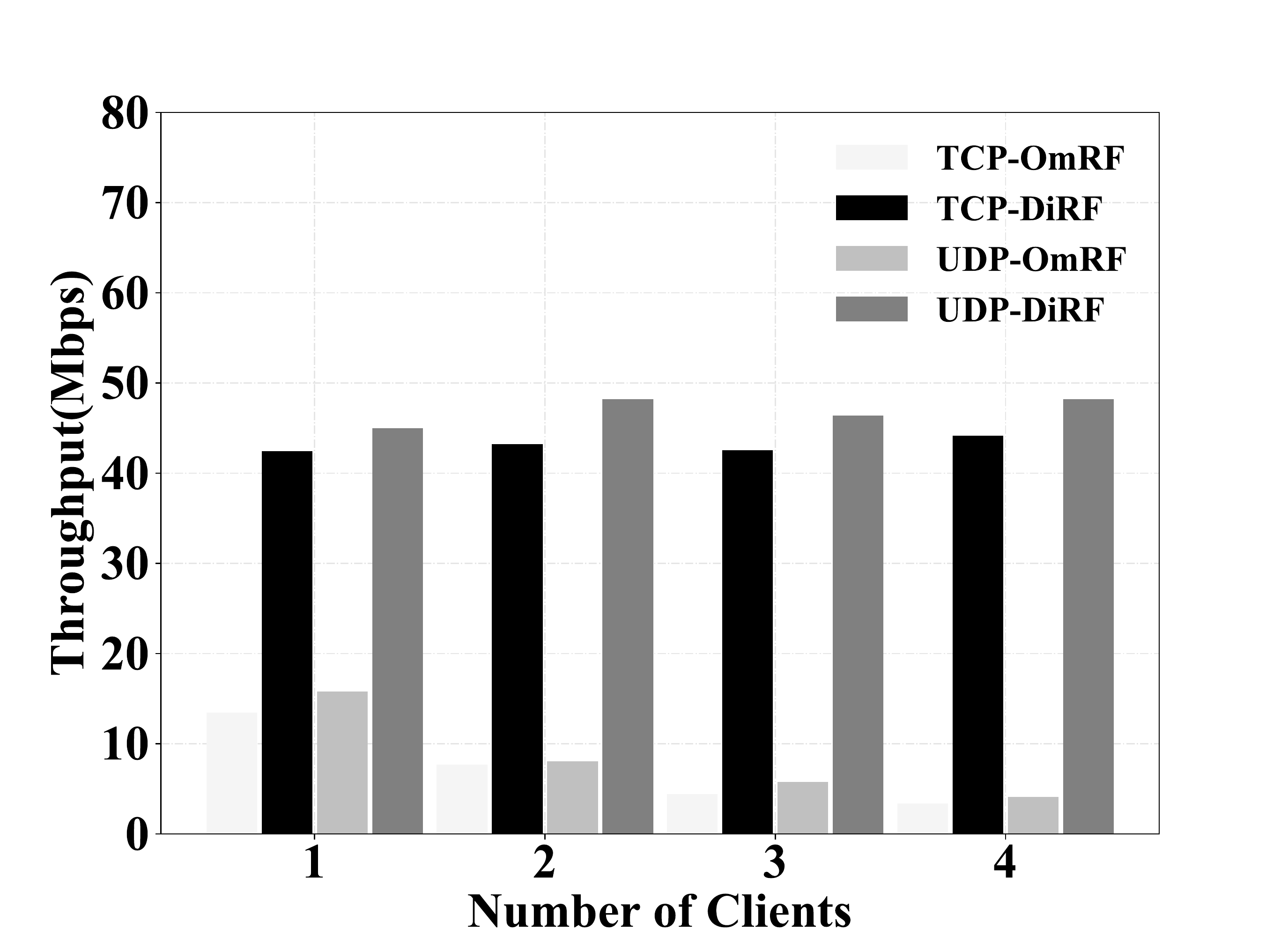}
  \caption{TCP and UDP Performance on 20MHz Bandwidth.}
  \label{subfig:multi_clients_20}
\end{subfigure}%
\begin{subfigure}{0.25\textwidth}
  \centering
  \includegraphics[width=1.0\linewidth, height=0.8\linewidth]{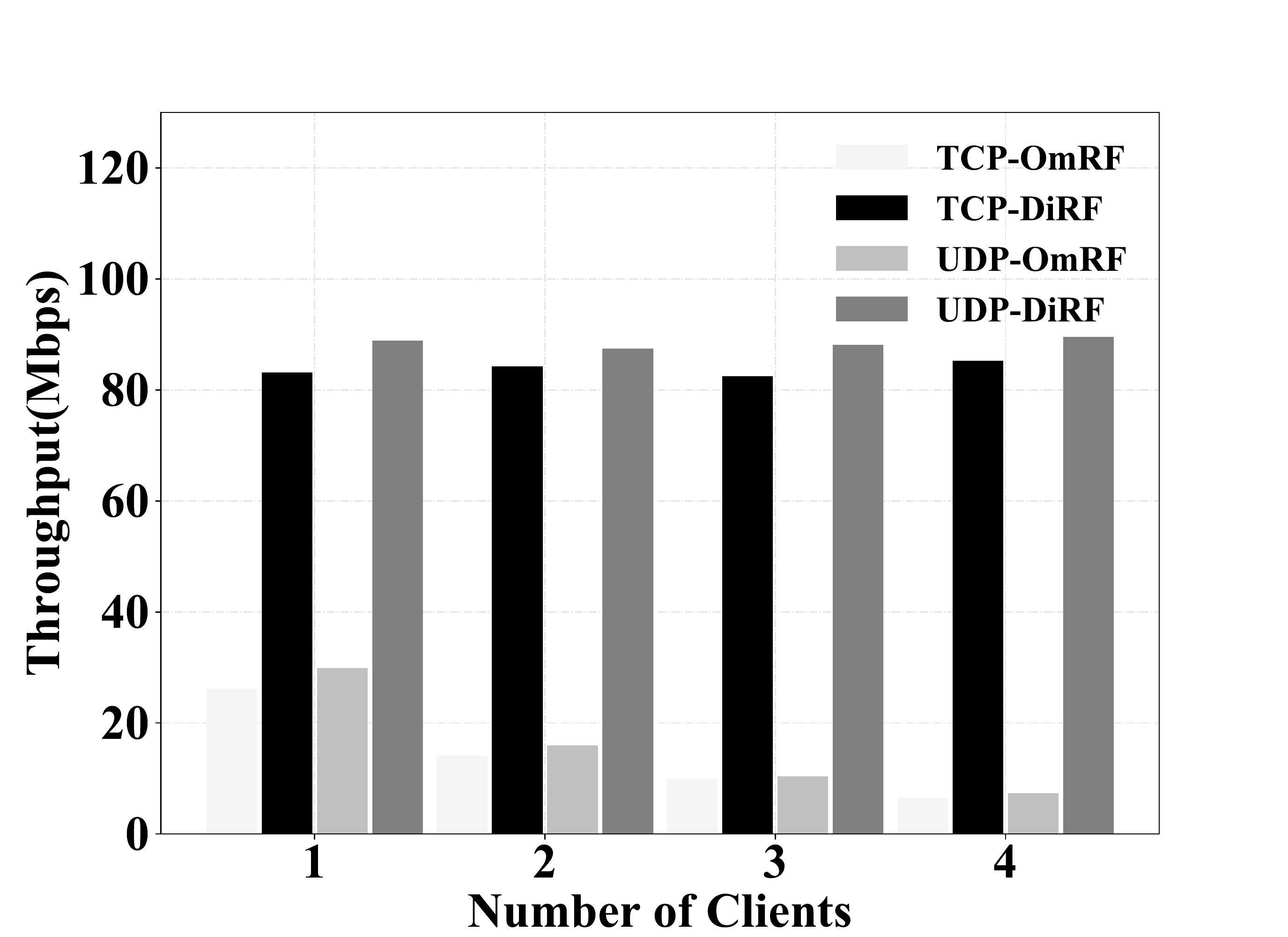}
  \caption{TCP and UDP Performance on 40MHz Bandwidth.}
  \label{subfig:multi_clients_40}
\end{subfigure}
\caption{TCP and UDP performance with different number of mobile clients.}
\label{fig:multi_clients}
\end{figure}

\textbf{Impact of the number of mobile clients.} We test \emph DiRF's performance, when there are multiple mobile clients moving around. In this experiment, we vary the number of mobile clients, and measure TCP and UDP throughput with 802.11 standard working at 5GHz on both 20MHz and 40MHz bandwidth. Fig.~\ref{fig:multi_clients} shows the results. In Fig.~\ref{fig:multi_clients}~(\subref{subfig:multi_clients_20}), we plot the TCP and UDP throughput, when working on 20MHz bandwidth. We can see that the average TCP and UDP throughput of \emph DiRF is $3.16\times$ and $2.85\times$ the average TCP and UDP throughput of \emph OmRF respectively, when there is one mobile client. When there are 4 mobile clients moving around, the average TCP and UDP throughput of \emph DiRF is $13.14\times$ and $11.76\times$ the average TCP and UDP throughput of \emph OmRF respectively. In Fig.~\ref{fig:multi_clients}~(\subref{subfig:multi_clients_40}), we plot the TCP and UDP throughput, when working on 40MHz bandwidth. We can see that the average TCP and UDP throughput of \emph DiRF is $3.18\times$ and $2.98\times$ the average TCP and UDP throughput of \emph OmRF respectively, when there is one mobile client. When there are 4 mobile clients moving around, the average TCP and UDP throughput of \emph DiRF is $13.11\times$ and $12.23\times$ the average TCP and UDP throughput of \emph OmRF respectively. As expected, \emph DiRF achieves high throughput, even though there are multiple mobile clients. However, both TCP and UDP throughput of \emph OmRF decreases as the number of the mobile clients increases. This is because the multiple mobile clients in \emph OmRF system have to carrier sense each other and compete the channel, which will degrade the throughput significantly. With \emph DiRF, different mobile clients can connect with different APs to have high throughput without channel competition.

\begin{figure}
\centering
\begin{subfigure}{0.25\textwidth}
  \centering
  \includegraphics[width=1.0\linewidth, height=0.8\linewidth]{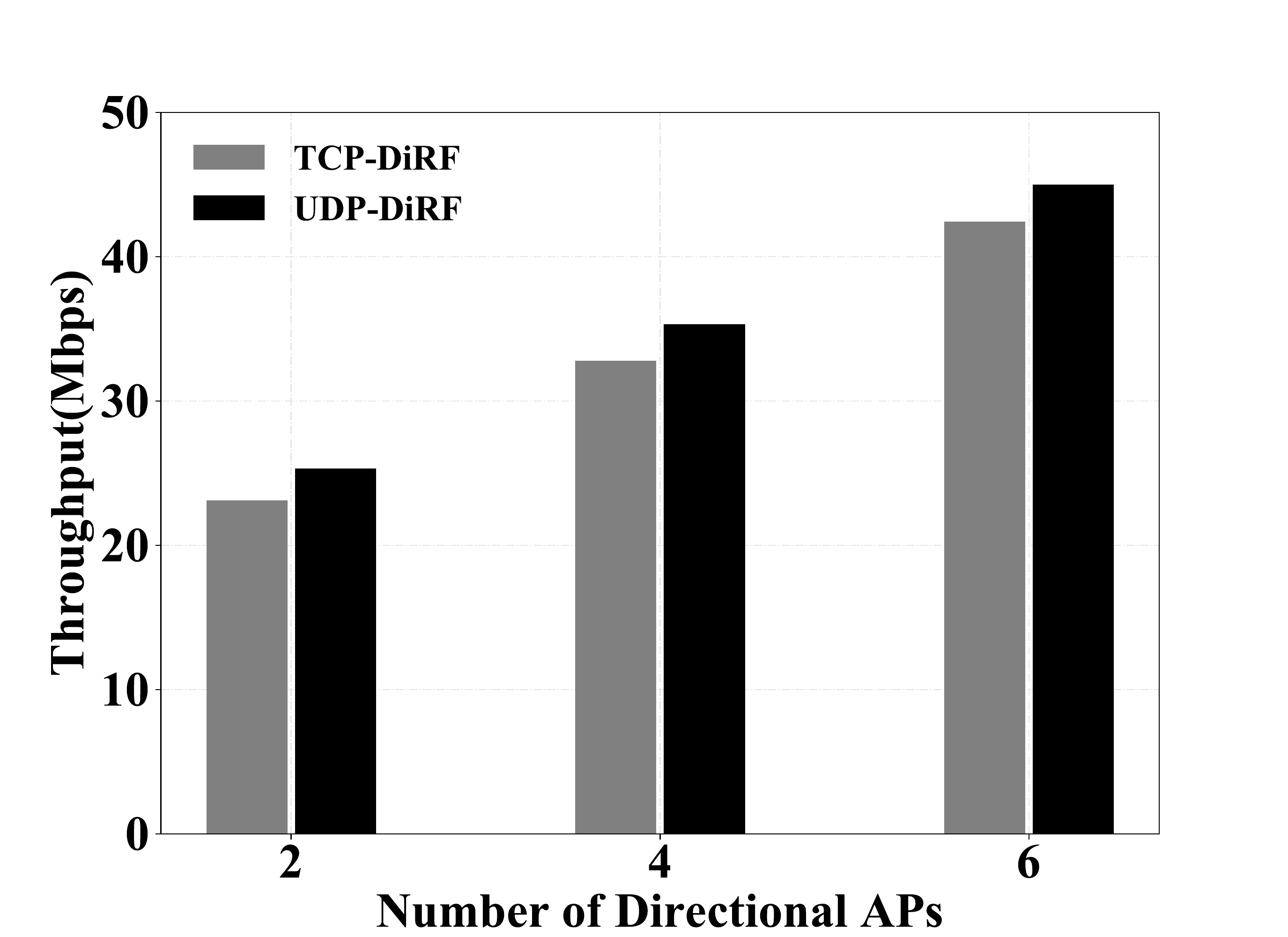}
  \caption{TCP and UDP Performance on 20MHz Bandwidth.}
  \label{subfig:multi_aps_20}
\end{subfigure}%
\begin{subfigure}{0.25\textwidth}
  \centering
  \includegraphics[width=1.0\linewidth, height=0.8\linewidth]{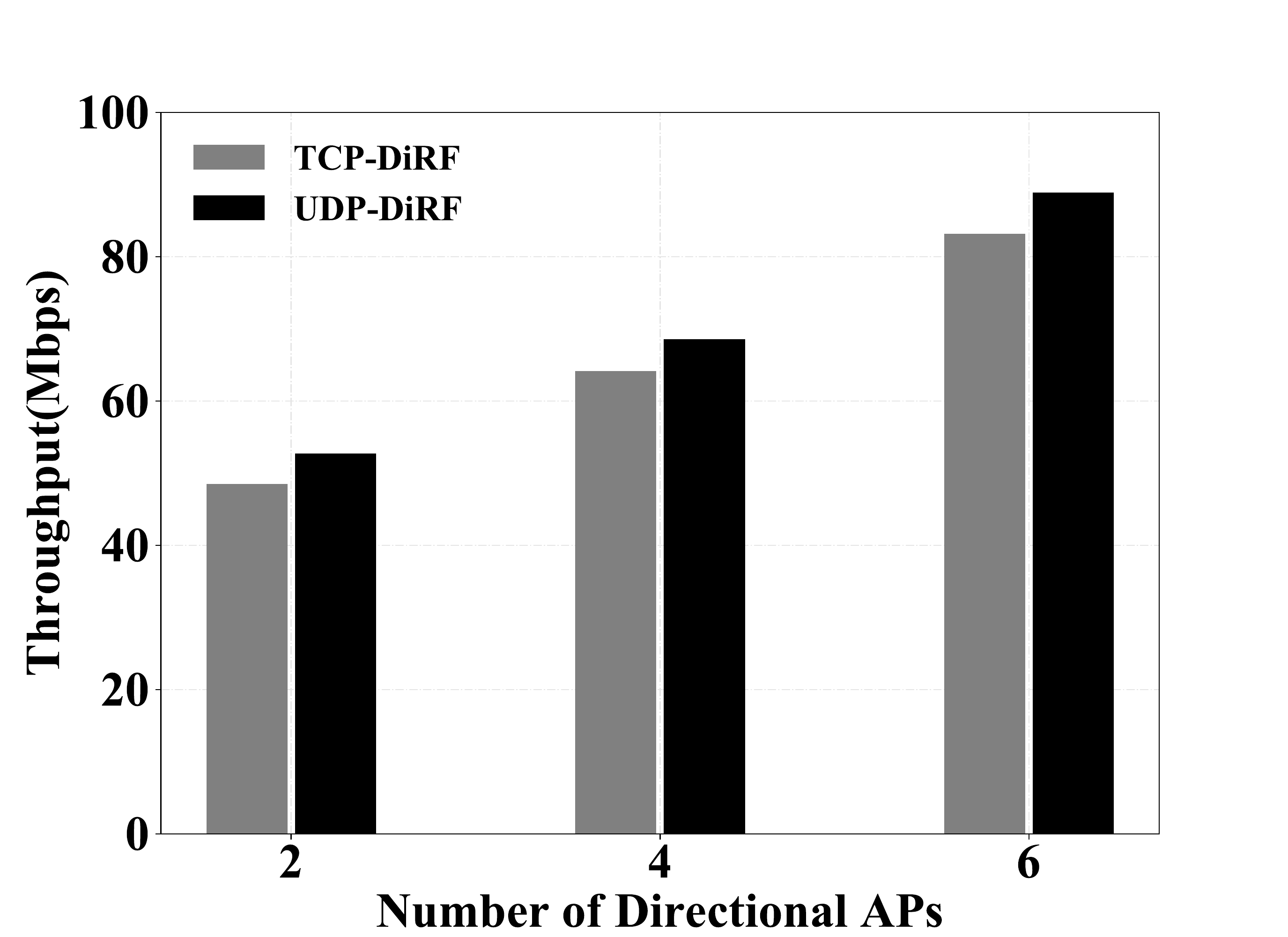}
  \caption{TCP and UDP Performance on 40MHz Bandwidth.}
  \label{subfig:multi_aps_40}
\end{subfigure}
\caption{TCP and UDP performance with different number of directional APs.}
\label{fig:multi_aps}
\end{figure}

\textbf{Impact of AP density.} We now examine the impact of AP density on the TCP and UDP throughput. In this experiment, we vary the number of APs mounted with the directional antennas, and measure the TCP and UDP throughput with 802.11 standard working at 5GHz on both 20MHz and 40MHz bandwidth. The results are shown in Fig.~\ref{fig:multi_aps}. In Fig.~\ref{fig:multi_aps}~(\subref{subfig:multi_aps_20}), we plot the TCP and UDP throughput, when working on 20MHz bandwidth. When there are two APs mounted with directional antennas in a room, the TCP and UDP throughput are 23.12Mbps and 25.31Mbps respectively. When there are six APs mounted with directional antennas in a room, the TCP and UDP throughput are 42.25Mbps and 45.01Mbps respectively. In Fig.~\ref{fig:multi_aps}~(\subref{subfig:multi_aps_40}), we plot the TCP and UDP throughput, when working on 40MHz bandwidth. When there are two APs mounted with the directional antennas in a room, the TCP and UDP throughput are 48.51Mbps and 52.71Mbps respectively. When there are six APs mounted with directional antennas in a room, the TCP and UDP throughput are 83.18Mbps and 88.89Mbps respectively. Hence, the TCP and UDP throughput increase as the number of the directional APs increases. This is because the directional APs can confine the communication in a narrow region of a room. More directional APs deployed in the room can cover larger place of a room, which will improve the spatial reuse to increase the throughput.

\subsection{Case Studies}

We now conduct two real-world case studies to test \emph DiRF's performance in the applications such as web downloads and web browsing.

\begin{figure}
    \centering
   \includegraphics[height=0.4\linewidth,width=0.7\linewidth]{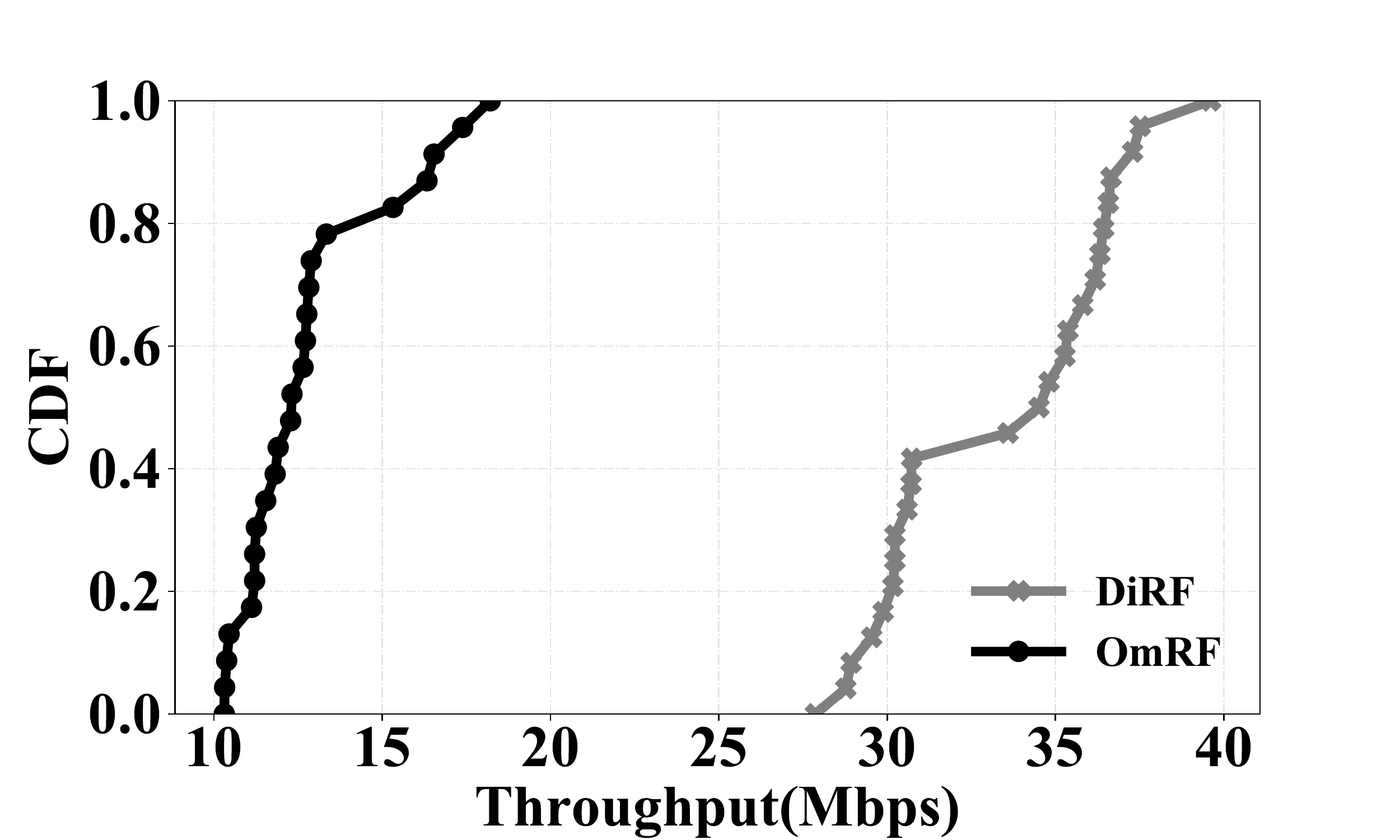}
        \caption{CDF of the throughput when downloading files from the web.}
        \label{fig:case_study}
\end{figure}

\textbf{Web downloads.} In this case study, we examine the performance of \emph DiRF and \emph OmRF on the file downloads from the web. The mobile client downloads a file with the size of 100MB during its movement in a room. To avoid the impact of Internet latency, the mobile client downloads the file directly from the local server. We use tcpdump to log the packets during downloading. Our experiments follow 802.11 standard working at 5GHz on 20MHz bandwidth. We measure the throughput of downlink packets on local server side. The result is shown in Fig.~\ref{fig:case_study}.  We plot the CDF of throughput of \emph DiRF and \emph OmRF. As shown, \emph DiRF can achieve 85\% percentile of 37.2Mbps. In contrast, \emph OmRF can achieve 85\% percentile of 16.2Mbps. And, the median throughput of \emph DiRF is $2.72\times$ the median throughput of \emph OmRF. This is because \emph DiRF can achieve high throughput communication during movement.

\textbf{Web browsing.} In this case study, we measure \emph DiRF's ability to load web pages for the mobile client. The mobile client browses a web page with the size of 2MB during its movement. We measure the time duration that the mobile client launches the chrome web browser until the web page is fully downloaded in its chrome web browser. To avoid the impact of Internet latency, the mobile client loads the web page from the local server directly. For both 20MHz and 40MHz bandwidth, we repeat the experiment 10 times, and the average loading time is shown in Table~\ref{tab:case_study}. As shown, the average loading time on 20MHz bandwidth with \emph OmRF is $2.84\times$ the average loading time on 20MHz bandwidth with \emph DiRF. And, the average loading time on 40MHz bandwidth is $2.86\times$ the average loading time on 40MHz bandwidth with \emph DiRF. This is because \emph DiRF can achieve high capacity communication for the mobile client during movement.

\begin{table}[t]
\caption{Web page loading time with different bandwidth.}
\begin{tabular*}{\linewidth}{@{\extracolsep{\fill}}lcr}
\hline
\bf Bandwidth  & \bf 20MHz & \bf 40MHz \\
\bf DiRF    & 0.37s & 0.22s \\
\bf OmRF  & 1.05s & 0.63s \\
\hline
\end{tabular*}
\label{tab:case_study}
\end{table}

\section{Related Work}
\label{section:related}

\textbf{Directional networks.} In comparison to omni-directional antenna, the directional antenna assembles the signals across all the spatial angles to form a directional beam, which can provide high data rate to the receiver in the transmitter's FoV. Therefore, there are many papers leveraging directional antennas to do indoor network capacity improvement~\cite{liu2009dirc, amiri2010directional,liu2010pushing} and outdoor network capacity improvement~\cite{navda2007mobisteer}. DIRC~\cite{liu2009dirc} is an indoor directional antenna system, which can improve the spatial reuse and network capacity in the indoor area. DIRC measures the optimal antenna orientation and uses a MAC protocol to maximize the transmission concurrency. Speed~\cite{liu2010pushing} is a distributed directional antenna system to maximize the spatial reuse with antenna configuration and control. And, antenna control contains antenna orientation algorithm, a MAC layer protocol and a novel client-AP association approach. However, the designed MAC protocol is not available in the commercial smartphones. MobiSteer~\cite{navda2007mobisteer} proposes a framework to maximize the throughput by steering the directional antennas in the outdoor area. However, MobiSteer selects the best AP based on the pretrained model, which cannot adapt to the highly random scenario. And, the overhead of steering the directional antenna is not unavoidable.  MiDAS~\cite{amiri2010directional} is a directional transmission system for mobile devices, which can select the right antenna for transmission. With rate adaptation and power control, MiDAS can further improve the goodput. The authors perform the measurement-driven experiments to show the link gain of passive directional antennas on the mobile devices. However, it takes further time to deploy based on the current network infrastructure.

\textbf{Millimeter wave networks and visible light communication.} Recently, millimeter wave (mmWave) networks and visible light communication (VLC) can provide high data rate. This is because they can concentrate the beam to one specific direction such that high throughput is provided due to high power gain. There are numerous papers trying to build a robust mmWave networks~\cite{wei2017pose, sur2017wifi, wei2017facilitating}. Pia~\cite{wei2017pose} leverages the mobile client's pose information to achieve seamless coverage in the 60GHz wireless network. Pia selects the best AP based on whether the mobile client and AP are in each other's FoV, which requires the statistical learning to predict AP's pose through mobile client's single walk in one AP's coverage. Also, there are some papers discussing the visible light communications~\cite{tian2016darklight, zhang2017pulsar,zhang2015extending}. DarkLight~\cite{tian2016darklight} can provide the sustained high throughput communication, when the LED appears dark or off. Okuli~\cite{zhang2015extending} and Pulsar~\cite{zhang2017pulsar} focus on localization by using visible light. However, mmWave networks and visible light communication are not widely deployed, because it needs further time and cost. After they are widely deplyed, it is important to use our \emph DiRF on top of them to improve the throughput during handoff.

\textbf{WiFi handover.} Numerous approaches are proposed to solve the AP selection problem during WiFi handover~\cite{song2017wi, croitoru2015towards, kandula2008fatvap, soroush2011concurrent, eriksson2008cabernet}. WiFi Goes to Town \cite{song2017wi} proposes a framework to achieve high throughput during handover with low cost WiFi chipsets and accurate CSI extracted from the commodity AP. WiFi Goes to Town leverages the effective SNR (ESNR)~\cite{halperin2011tool} extracted from the AP to make a decision on which AP the mobile client connects with. And, block acknowledgement mechanism is suggested to avoid the retransmissions after AP switching. However, the selected AP with optimal ESNR is not the optimal selection in the near future without considering the mobile client's moving direction. In~\cite{croitoru2015towards}, the authors leverage the multipath TCP (MPTCP) to concurrently connect with multiple APs with different virtual network interfaces in order to have high throughput during WiFi handover. However, MPTCP is not widely deployed and the physical layer performance also restricts the high layer throughput. Spider~\cite{soroush2011concurrent} predicts the vehicular client's mobility and connectivity pattern, and allows it to associate with multiple APs working on different channels. In contrast to Spider, \emph DiRF works in indoor area and needs more accurate clients' positions.

\section{Discussion and Future Work}
\label{section:discussion} 

\textbf{Interference management.} First, our current prototype does not consider interference with the nearby APs working on the same channel. We can let the adjacent APs work on different wireless channel to avoid the interference and improve the network throughput of our system. However, the spectrum efficiency would be degradation. Moreover, the channel switching latency is an overhead. Second, our prototype does not consider the non-line-of-sight (NLOS) interference caused by the environmental reflectors. We just consider the line-of-sight (LOS) path. This is because when we use the high-gain directional antenna, the LOS path is the dominant path as the directional antenna is mounted on the ceiling of the room. With high-gain directional antenna, our system is robust to the environmental reflectors. Also, we can use statistical learning to model the main environmental reflectors and mitigate the NLOS interference afterwards.

\textbf{Position estimation.} Our current prototype just considers 3-DoF information to estimate the mobile client's position and AP's position. When the mobile client has the directional antenna, we can leverage 5-DoF information including 3D location, polar and azimuth angle to improve AP selection. And, we can consider angle of arrival (AoA) and angle of departure (AoD) estimation, which can be measured by the phased array antenna. With AoA and AoD, we can improve the position estimation and network throughput.

\section{Conclusion}
\label{section:conclusion}
We have presented \emph DiRF, which is the high-gain directional antenna based indoor wireless access network designed to achieve high capacity communication in the indoor area with densely deployed directional APs. The directional antenna mounted on the ceiling of the room provides high capacity communication. And, we implement position based AP selection algorithm to achieve high capacity communication during AP switching and downlink packet scheduler to decrease the retransmissions caused by AP switching.
\balance

\bibliographystyle{IEEEtran}
\bibliography{sample-bibliography}

\end{document}